\documentstyle[prd,aps]{revtex}

\begin{document}
\draft

\title{Search for periodic gravitational radiation with the ALLEGRO gravitational wave detector}
\author{E.~Mauceli\footnote{Present address: Center for High Energy Physics, University of Oregon, Eugene,
OR, 97403, USA}, M.~P.~McHugh, W.~O.~Hamilton, W.~W.~Johnson, and
A.~Morse }
\address{Department of Physics and Astronomy, Louisiana State University, Baton Rouge, LA
70803, USA}   

\date{\today}
\maketitle

\begin{abstract} We describe the search for a continuous signal of gravitational radiation from
a rotating neutron star in the data taken by the ALLEGRO gravitational wave detector in early
1994.  Since ALLEGRO is sensitive at frequencies near 1 kHz, only neutron stars with spin
periods near 2 ms are potential sources.  There are no known sources of this type for ALLEGRO,
so we directed the search towards both the galactic center and the globular cluster 47
Tucanae.  The analysis puts a constraint of roughly $8 \times 10^{-24}$ at
frequencies near 1 kHz on the gravitational strain emitted from pulsar spin-down in either 47
Tucanae or the galactic center. 

\end{abstract}


\section{Introduction}
\label{sec:intro}

The majority of experimental searches for gravitational radiation have focused on the detection
of burst signals, such as those emitted from the collapse of a massive star.  There are
compelling arguments that nearby millisecond pulsars can provide a detectable source of
continuous (CW) gravity waves~\cite{new_apj95}-~\cite{owen_prd98}. We will use the term ``pulsar''
in the following text to refer to a rotating neutron star, regardless of whether it is  emitting
detectable electromagnetic radiation.

It is well known that a symmetric object rotating about its symmetry axis does not emit
gravitational radiation.  Therefore, if a pulsar is to radiate gravity waves, it must either be
asymmetric or precessing.  For the purposes of this search, we have considered only asymmetric
pulsars.  The asymmetry can be rotationally induced or due to a ``star quake'' deforming the
neutron star crust or due to some other process. By any mechanism, the amount of ellipticity
produced in the neutron star is expected to be small, from a maximum of
$10^{-5}$, which, for some pulsar models is the maximum supportable asymmetry of the neutron
star crust~\cite{shapiro_83} to less than $10^{-9}$.  The latter number is obtained by assuming
the measured spin-down of certain millisecond pulsars is due entirely to the emission of
gravitational radiation.

In this article we report the results of our first search for a CW signal in data taken by the
ALLEGRO resonant gravitational wave detector at Louisiana State University during the first
three months of 1994~\cite{mauceli_prd96}.  Another search has been performed by the EXPLORER
detector team from the University of Rome, which has a similar detector and is pursuing a
different type of analysis~\cite{astone_96}.

If there were known pulsars with the right spin period, they would be the natural target for
any CW search. But there are only a small number of known pulsars with spin periods small
enough (2.2 milliseconds) to match our antenna's 900 Hz frequency 
and none of those listed in the 1995 Taylor pulsar catalog~\cite{taylor_95} has a spin period
that matches our two narrow reception bands.  We would ideally like to make an all sky
search, but this is computationally very intensive.  Given limited computing resources
it nearly impossible to exhaust all of the obvious possibilities.  All-sky search strategies
have been proposed elsewhere~\cite{schutz_papa}. 

Therefore we have chosen another strategy and directed our first search towards two regions
where one might reasonably expect to find a high density of CW sources. The first is the
globular cluster 47 Tucanae (Tuc), and the second is the center of the galaxy.  The celestial
coordinates for both are listed in Table~\ref{tab:coords}.  We chose 47 Tuc because it is
relatively close, and because of the large number of fast millisecond pulsars known to be
concentrated in this globular cluster~\cite{parkespeople}, which suggests the possibility of
more undetected neutron stars with short spin periods.   

The analysis assumes the simplest possible CW source, one that has a stable emission frequency
in its rest frame. In other words it is not spinning down and has no companions to cause
orbital doppler shifts.  In this case only knowledge of the earth's motion and the antenna's
reception pattern are needed to determine the modulation of the frequency and amplitude of the
received signal. These modulations are highly dependent on the source location.  At the
frequency resolution of our search, if the actual source location is offset from the supposed
location by as little as 1.00$'$ in right ascension and 0.25$'$ in declination, the
gravitational wave frequency will be mis-identified by 1 bin~\cite{mauceli_phd97}.  However,
the range of possible locations for detection of the signal (albeit with the frequency mis-identified)
is significantly larger~\cite{brady_prd97}.

\section{ALLEGRO}
\label{IA}

The ALLEGRO gravitational wave detector~\cite{mauceli_prd96} is located in the Physics Building
at Louisiana State University in Baton Rouge, Louisiana ($30^{\circ}.413$ N lat, $
91^{\circ}.179$ W long).  ALLEGRO consists of a resonant aluminum bar equipped with a resonant
inductive transducer and a dc SQUID amplifier all cooled to 4.2 K.   The bar is 60 cm in
diameter and 300 cm in length, with a physical mass of 2296 kg.  The bar is oriented
perpendicular to the plane of the great circle on the Earth that passes through Geneva, the
location of the Rome Explorer antenna, and midway between Baton Rouge, LA and Stanford, CA. 
This orientation results in the axis of ALLEGRO being directed along a line
$40^{\circ}.4$ west of North.  

Vibrations of our resonant antenna produce a voltage out of the SQUID electronics which is
proportional to the relative displacement of the bar and transducer.  The coupled system of bar
and transducer has two normal modes at roughly 896.8 Hz and 920.3 Hz.  These are referred to as
the ``minus'' and ``plus'' resonant modes respectively.  ALLEGRO is most sensitive in a small
bandwidth around each of these modes.  The voltage out of the SQUID electronics is sent to a
single lock-in detector which demodulates and low pass filters the signal.  The reference
frequency of the lock-in is set halfway between the normal mode frequencies of the antenna,
thus shifting the frequency of the signal from near 1 kHz to near 10 Hz.  The output
consists of in-phase ($x$) and quadrature ($y$) components which contain the full amplitude
and phase information.  The demodulated data is then sampled 125 times a second and written
to disk. 

\section{target source}
\label{sec:source}

We describe the gravity wave in a reference frame which has its $\hat{e}_{z^{'}}$ axis aligned
with the direction of propagation of the wave.  This frame is referred to as the ``wave frame''
and is denoted by primed coordinates ($ x^{'},y^{'},z^{'}$).  The most general form of the
gravity wave in the wave frame is 
\begin{equation} 
  {\mathbf h}_{wave} =  
\left[
     \begin{array}{ccr} 
              h_{+}(t^{'}) & h_{\times}(t^{'}) & 0 \\
               h_{\times}(t^{'}) & -h_{+}(t^{'}) & 0 \\
               0 & 0 & 0  
     \end{array} \right]  
\label{h}
\end{equation}   where $h_{+}(t^{'})$ and $h_{\times}(t^{'})$ are the amplitudes of the two
allowed states of linear polarization, referred to as the plus and cross amplitudes
respectively.

In General Relativity one can compute the polarization amplitudes for a rotating, asymmetric
neutron star using the quadrupole approximation~\cite{MTW}. We choose for our model a neutron
star with the three principle moments of inertia about  the three principle axes fixed in the
body frame, denoted by $I_{1},I_{2} \, \mbox{and}\, I_{3}$.  The rotation axis is chosen to be
along $I_{3}$ and the neutron star is assumed to be deformed such that $I_{1} \neq I_{2}$ . 
The ellipticity of the pulsar is then defined to be
\begin{equation} 
  \epsilon = \frac{I_{1}-I_{2}}{I_{3}} \, .
  \label{asymm}
\end{equation}  The resulting gravitational radiation is emitted at twice the pulsar rotation
frequency and the two polarization amplitudes are given by
\begin{eqnarray} \label{hcross}
  h_{+}(t^{'}) &=& h_{c}\, (1+\cos^{2} i ) \, \cos (2 \pi f_{0} t^{'}) \\ 
  h_{\times}(t^{'}) &=& 2\,h_{c} \cos i \, \sin (2 \pi f_{0} t^{'}) \nonumber
 \end{eqnarray}
 where the angle $i$ is measured between the pulsar spin axis and the line of
sight to the detector, $h_{c}$ is the ''characteristic amplitude''~\cite{new_apj95} of the
incident wave, and $f_{0}=2 f_{rot}$ is the frequency of the gravitational wave, equal to twice
the rotation frequency of the pulsar.  This is not the most general pulsar model one could
construct. If the neutron star deformation is not along a principle axis, then emission occurs
at the rotation frequency and twice the rotation frequency~\cite{galtsov_jetp84}.   If the star
is precessing, emission occurs at the rotation frequency, and the rotation frequency plus or
minus the precession frequency~\cite{zimmermann_prd79}.   For this work, we consider only
emission at twice the rotation frequency.

The characteristic amplitude\footnote{New's characteristic strain is greater by a factor of 2
as they assumed that the rotation axis of the pulsar was along the line of sight to the Earth.}
is given by
\begin{equation} 
  h_{c} = \frac{2G}{c^{4}r}\epsilon I_{3}\, (\pi f_{0})^{2}
  \label{hc}
\end{equation} where $r$ is the distance to the source, $G$ is the gravitational constant, and
$c$ the speed of light.  Adopting a value of $I_{3}=10^{45}\, \mathrm{g \, cm}^{2}$ (a
reasonable estimate for a $1.4 M_{\odot}$ pulsar with a radius of 10 km), this can be written as
\begin{equation}
   h_{c} = 5.2\times 10^{-28} (\frac{I_{3}}{10^{45} \,\mathrm{g \, cm}^{2}})\,
         (\frac{\epsilon}
       {10^{-8}})\, (\frac{10 \, \mathrm{kpc}}{r})\,  (\frac{f_{0}}{1 \, \mathrm{kHz}})^{2} \, .
\label{hcdimen}
\end{equation} For galactic sources ($<$ 10 kpc), and assuming the maximum allowed pulsar
ellipticity ($\epsilon \sim 10^{-5}$) we see that the characteristic amplitude is of order
$10^{-24}$.  As will be shown, this is of the same level as the measurements.

\section{phase considerations}
\label{sec:dopshift}

Reliable detection of a periodic gravitational wave signal depends on tracking the phase of the
signal through many cycles.  As given in Eq.~\ref{hcross}, the polarization amplitudes of the
gravity wave are pure sinusoids at an unknown frequency.  Detection of such a signal would
involve a single Fourier transform of the data, the resulting spectrum then being scanned for
anomalous peaks.  However, the phase of the gravity wave as observed at the detector at a
particular observation time, relative to the emitted phase, depends on a number of factors: 1)
intrinsic pulsar spin-down, 2) motion of the Earth within the solar system, 3) if it is part
of a binary system, orbital motion of the pulsar, and 4) proper motion of the pulsar.  This
results in the initially narrow-band signal being spread out over many frequency bins,
decreasing the signal to noise ratio of the single Fourier transform.  As the expected
signals are already weak, this seriously compromises one's detection capabilities.  For the
purposes of this work, we assume that the pulsar is solitary with negligible spin-down, and
has no other accelerations with respect to the solar system barycenter (SSB).  We also assume the 
proper motion of the pulsar is small such that it does not move significantly during the observation
time.   

It should be
noted that the spindown due to the emission of gravitational radiation at the level of
sensitivity of this experiment would be quite significant. The lack of an actual target
source has led us to choose the simplest case for the purposes of demonstration of the potential
sensitvity of a real detector. With these assumptions, the phase of the carrier frequency is
modulated only by motion of the Earth in the solar system, with the modulation being given by 
(see Fig.~\ref{fshift})
\begin{equation}
\Phi(t) = \omega_{0} (\, \frac{1}{c} {\mathbf r}(t) \cdot \hat{{\mathbf n}} +
1.658^{ms} \sin g(t) \, )
\label{dopshift}
\end{equation}
where ${\mathbf r}(t)$ is the vector
from the SSB to the center of mass of the detector at observation time $t$,
$\hat{{\mathbf n}}$ is a unit vector from the SSB directed towards the pulsar, $c$ is the
speed of light, and $\omega_{0} = 2 \pi f_{0}$.  The second term is due to the combined effect of
time dilation and gravitational redshift due to the solar system bodies.  It is periodic over a
year with maximum delay of 1.7 ms.  We note that the angle $g(t)$ varies slightly from year to
year.  In 1994 its value was 
\[ g(t)/2\pi = 356.60 + 0.98560 \times D \] 
where $D$ is the day of the year.  Including this information into Eq.~\ref{hcross}, we
write the two polarizations of the gravity wave as (expressed in a frame parallel to the
wave frame)
\begin{eqnarray} \label{hphase}
  h_{+}(t) &=& h_{c}\, (1+\cos^{2} i ) \, \cos (\omega_{0} t + \Phi(t) + \Phi_{0}) \\ 
  h_{\times}(t) &=& h_{c}\, (2 \cos i) \, \sin (\omega_{0} t + \Phi(t) + \Phi_{0})
\nonumber
\end{eqnarray}
where $\Phi_{0}$ is an unknown, constant offset.

Astronomical locations are usually given as right ascention and declination coordinates,
denoted $\alpha,\delta$ respectively.  These are defined in the the ``celestial'' coordinate 
frame (CC frame with coordinates $(X,Y,Z)$).  This frame is centered at the solar system
barycenter (SSB) with $\hat{e}_{Z}$ along the Earth's rotation axis.
$ \hat{e}_{X}$ and $\hat{e}_{Y}$ are in the Earth's equatorial plane with 
$ \hat{e}_{X}$ directed towards the intersection of the equatorial plane with the Earth's
orbital plane (the ecliptic) at the vernal equinox.  Right ascension is measured in hours of
angle (12 hrs = $\pi$ radians) from the vernal equinox eastward along the celestial equator to
the celestial object and declination is measured in degrees north (+) or south (-) of the
equatorial plane.  

To calculate the dot product in the phase modulation term, we also expressed the detector
position in the CC frame.  This was done in two steps.  First, the position of the  center of
mass of the Earth relative to the SSB was obtained using a commercially available software
package from the U.S. Naval Observatory (MICA\footnote{Multiyear Interactive Computer Almanac,
U.S. Naval Observatory, 3450 Mass. Ave., N.W. Washington, DC 20392}).  MICA uses the Jet
Propulsion Laboratory DE200/LE200 ephemeris.  It reports positions of the Earth's center of
mass in a cartesian coordinate frame centered at the SSB.  Coordinates are given to the nearest
$10^{-9}$ astronomical units (AU), which is of order $10^{2}$ meters.   This corresponds to
1/1000 of a wavelength for the signals of interest here, enabling an accurate tracking of the
phase of the gravitational wave.  

Second, a GPS receiver was used to obtain the latitude and longitude coordinates of ALLEGRO, 
again with sufficient accuracy to track the signal.  These were converted to cartesian
coordinates in a frame centered at the Earth's center of mass and vectorally added to the
center of mass positions to provide ALLEGRO's position with respect to the SSB.  The cartesian
coordinates for ALLEGRO were converted to the spherical coordinates ($r,t_{s},\delta_{A}$):
$r$ being the distance from the SSB to the detector, $t_{s}$ the local sidereal time, and
$\delta_{A}$ the declination of the detector.  Using the spherical trigonometric formula for
the cosine of the angle between two vectors, we write the phase delay as
\begin{equation}
\Phi(t) = 2\pi f_{0} \left (\, \frac{r(t)}{c} [\sin \delta_{A}\, \sin \delta + \cos
\delta_{A}\, \cos \delta\, \cos\gamma(t) \,] + 1.658^{ms} \sin g(t) \,\right )
\label{phse2}
\end{equation} 
where $\gamma = t_{s}-\alpha$ is the local hour angle.

\section{signal considerations}
\label{sigcon}

A passing gravity wave provides the largest fractional change in the length of a bar, and
therefore the largest signal in a bar detector, when its direction of propagation is
perpendicular to the bar axis and one of the polarizations of the gravity wave lies along the
bar axis.  In general, the gravity wave is incident to the bar with some arbitrary angle and
polarization.  We define the polarization angle as the angle between the bar axis (west of
North) and the direction of the plus polarization of the gravity wave.  

We describe the detector in a coordinate frame whose origin is at the center of mass of the
antenna, with the $\hat{e}_{z}$ axis aligned with the local vertical and the
$\hat{e}_{x}$ axis aligned with the bar axis. This frame is referred to as the ``lab frame'' 
and is denoted denoted by unprimed coordinates ($x,y,z$).  

For a gravity wave incident with an arbitrary orientation to the bar axis, it is only the
component of the strain tensor along the bar axis which produces a detectable driving force on
the bar.  This force is commonly written as
\begin{equation}
   {\cal F}(t) = \frac{1}{2} \mu L_{e} \ddot{h}_{xx}(t) 
\label{force}
\end{equation}              
where $\ddot{h}_{xx}(t)$ is the second time derivative of the
strain component along the bar axis.  The quantities $\mu$ and $L_{e}$ are the effective mass
and length of the first longitudinal eigenmode of the bar, obtained by solving the elastic
equations of motion for a long, thin cylinder.  The effective mass is equal to half the
physical mass of the bar ($\mu=1148$ kg) and $L_{e} = (4/\pi^{2})$L where L=3 m is the actual
length of the bar.  The force acting on the bar dominates the output as the force on the
transducer itself produces a much smaller motion.

To calculate the component of the gravitational wave, given by Eq.~\ref{h} in the wave frame,
along the bar axis requires knowledge of the rotation matrix between the wave frame and the lab
frame.  As both source direction and detector locations are known, the rotation matrix is
completely specified (up to an angle related to the unknown polarization of the gravity wave). 
We calculate the rotation matrix by first rotating the wave frame to the CC frame, and then the
CC frame to the lab frame.  The full rotation matrix from the wave frame to the lab frame is
given by
\begin{equation}
  {\mathbf R}_{wave\rightarrow lab} = {\mathbf R}_{CC\rightarrow lab} 
   {\mathbf R}_{wave\rightarrow CC} \, .
\label{fullrot}
\end{equation}

The CC frame is related to the wave frame by the angles
$(\phi=\alpha-\pi/2,\theta=\delta+\pi/2,\psi=\psi_{0})$, where we have used the Euler
x-convention to define the axes of rotation.  Using Eq.~4-46 of Goldstein~\cite{goldstein_80},
with the stated angular substitutions, the rotation matrix from the wave frame to the CC frame
is given by
\[ {\mathbf R}_{wave\rightarrow CC} = \left( \begin{array}{ccc}
 (\sin \alpha \cos \psi_{0} - \cos\alpha \sin\delta\sin\psi_{0}) &
 (-\sin\psi_{0} \sin\alpha - \cos\psi_{0} \cos \alpha \sin \delta) &  (-\cos \alpha  
\cos \delta )\\
(-\cos\psi_{0} \cos \alpha -\sin\psi_{0} \sin\alpha \sin\delta )& (\sin\psi_{0} \cos\alpha 
- \cos\psi_{0} \sin
\alpha \sin \delta) & (-\sin\alpha \cos\delta) \\ (\sin\psi_{0} \cos \delta) &( \cos\psi_{0} 
\cos\delta
)& (-\sin\delta)
\end{array} \right) \, .
\] 
Again using the x-convention to define the rotation axes, the CC frame is related to the
laboratory frame by the angles
$(\phi=t_{s}+\pi/2,\theta=\pi/2-l,\psi=\pi/2+\eta)$.  $t_{s}$ is the detector local sidereal
time as before and $l$ is the detector latitude.  The particular value for the final rotation
angle comes from the choice to define the $\hat{e}_{x}$ axis pointing west of North along the
bar axis.  The rotation matrix from CC frame to lab frame is then
\[ 
 {\mathbf R}_{CC\rightarrow lab}=\left( \begin{array}{ccc} (\sin\eta \sin t_{s} - \sin l \cos
t_{s} \cos\eta) & (-\sin\eta \cos t_{s} -\sin l \sin t_{s}
\cos\eta) & (\cos\eta \cos l) \\  (\cos\eta \sin t_{s} + \sin l \cos t_{s} \sin\eta)
&(-\cos\eta \cos t_{s}+\sin l \sin t_{s}
\sin\eta)  & (-\sin\eta \cos l )\\ (\cos l \cos t_{s}) & (\cos l \sin t_{s}) & (\sin l)
\end{array} \right) \: .
\] 
Using Eq.~\ref{fullrot}, the full rotation matrix may be computed and $h_{xx}$ is the (11)
component of
\begin{equation}
  {\mathbf h}_{lab}  = {\mathbf R}_{wave \rightarrow lab} {\mathbf h}_{wave} 
  {\mathbf R}^{T}_{wave \rightarrow lab}\, .
\label{hinlab}
\end{equation}
Considering only the component of the gravity wave along the bar axis,
\[ 
h_{bar}(t) = (R^{2}_{11}-R^{2}_{12})\, h_{+}(t) + 2\, R_{11}\, R_{12}\, h_{\times}(t) \, 
\]
where the relevant components of the rotation matrix are
\[ 
\begin{array}{llc} 
R_{11} & = \cos\psi_{0} (\sin\eta \cos\gamma + \sin l \cos\eta \sin\gamma) 
\\ & + \sin\psi_{0} (-\sin\eta
\sin\delta \sin\gamma + \sin l \cos\eta \sin\delta \cos\gamma + \cos\eta \cos l \cos\delta) 
       \\ 
R_{12} &= \cos\psi_{0} (-\sin\eta \sin\delta \sin\gamma + \sin l \cos\eta
  \sin\delta \cos\gamma + \cos\eta \cos l \cos\delta) \\ & - \sin\psi_{0} (\sin\eta \cos\gamma
 + \sin l \cos\eta \sin\gamma) \, .\\
\end{array} 
\]
Defining
\begin{eqnarray}\label{fp} 
f_{+}(t,\eta,\gamma,\delta, l) &=& (\sin\eta \cos\gamma(t) + \sin l \cos\eta
\sin\gamma(t))^{2} 
\\
& &- (-\sin\eta \sin\delta \sin\gamma(t) + \sin l \cos\eta \sin\delta \cos\gamma(t) + 
  \cos\eta \cos l \cos\delta)^{2} \nonumber
\end{eqnarray}
and
\begin{eqnarray}\label{fm} 
f_{\times}(t,\eta,\gamma,\delta, l) &=& 2\, (\sin\eta \cos\gamma(t)
+ \sin l \cos\eta \sin\gamma(t))
 \\ 
 & & \times(-\sin\eta \sin\delta \sin\gamma(t) + \sin l \cos\eta \sin\delta \cos\gamma(t) + 
\cos\eta \cos l \cos\delta) \nonumber 
\end{eqnarray} 
we have 
\begin{eqnarray} 
R^{2}_{11}-R^{2}_{12} &=& \cos 2 \psi_{0}\, f_{+}(t) + 
\sin 2 \psi_{0}\, f_{\times}(t) \nonumber \\ 
2\, R_{11}\, R_{12} &=& \cos 2 \psi_{0}\, f_{\times}(t) - \sin 2 \psi_{0}\, f_{+}(t) \nonumber
\end{eqnarray}
and
\begin{equation} 
h_{bar}(t) = [ \cos 2 \psi_{0}\, f_{+}(t) + \sin 2 \psi_{0}\, f_{\times}(t) ] \, h_{+}(t) + 
[\cos 2 \psi_{0}\, f_{\times}(t) - \sin 2 \psi_{0}\, f_{+}(t) ] \, h_{\times}(t) \, .
\label{hang}
\end{equation}
Using Eq.~\ref{hphase} and Eq.~\ref{hang}, the waveform is given by
\begin{eqnarray}
\label{waveform} 
h_{bar}(t) &=&  h_{c}\,(1+\cos^{2} i ) \, 
[ \cos 2 \psi_{0} \, f_{+}(t) + \sin 2 \psi_{0}\, f_{\times}(t) ]\, \cos (\, \omega_{0} t + 
\Phi(t) + \Phi_{0} \,) \\  
&+& h_{c}\, (2\cos i) \, 
[\cos 2 \psi_{0} \, f_{\times}(t) - \sin 2 \psi_{0} \, f_{+}(t) ]\, \sin (\, \omega_{0} t + 
\Phi(t) + \Phi_{0}\, ) \, .
\nonumber 
\end{eqnarray}
The reception patterns $f_{+}(t)$ and $f_{\times}(t)$ are shown in Fig.~\ref{recpat} for a signal
from 47 Tuc.
Writing the time-dependent sine and cosine terms in Eq.~\ref{waveform} as exponentials
and defining
\begin{eqnarray}
  F_{+}(t) = \exp [j \Phi(t) ] f_{+}(t) \\ 
  F_{\times}(t) = \exp [j \Phi(t) ] f_{\times}(t) 
  \nonumber 
\end{eqnarray} 
the expression for the gravity wave strain along the bar axis is
\begin{eqnarray}
\label{waveform3} 
h_{bar}(t) &=& \frac{1}{2} h_{c}\,(1+\cos^{2} i )\, \exp (j \Phi_{0}) \,
[ \cos 2 \psi_{0} F_{+}(t) + \sin 2 \psi_{0} F_{\times}(t)] \, \exp (j \omega_{0} t) \\
&+& \frac{1}{2} h_{c}\,(1+\cos^{2} i )\, \exp (-j \Phi_{0}) \,
[ \cos 2 \psi_{0} F^{*}_{+}(t) + \sin 2 \psi_{0} F^{*}_{\times}(t)] \,\exp (-j \omega_{0} t) 
\nonumber \\
&-& \frac{j}{2}\, h_{c}\, (2\cos i) \, \exp (j \Phi_{0}) \,
[\cos 2 \psi_{0} \, F_{\times}(t) - \sin 2 \psi_{0} \, F_{+}(t) ]\,\exp (j \omega_{0} t) 
\nonumber \\
&+& \frac{j}{2}\, h_{c}\, (2\cos i) \, \exp (-j \Phi_{0}) \, 
[\cos 2 \psi_{0} \, F^{*}_{\times}(t) - \sin 2 \psi_{0} \, F^{*}_{+}(t) ] \,
\exp (-j \omega_{0} t) \nonumber \, .
\end{eqnarray}
%
%
%
We are now in position to calculate the anticipated signal as it would appear in the ALLEGRO
data stream.  This is most clearly presented in the frequency domain. 
Fourier transforming Eq.~\ref{force}, the force produced on the bar is
\begin{equation}\label{dforce}
 {\cal F}(\omega) = -\frac{1}{2} \mu L_{e}\, \omega^{2}\, h_{bar}(\omega) \, .
\end{equation} 
This driving force produces motion of the transducer 
\[ H(\omega) = G(\omega)\, {\cal F}(\omega) = -\frac{1}{2} \mu L_{e}\, \omega^{2}\, G(\omega) \, 
h_{bar}(\omega)
\]  
where $G(\omega)$ is the transfer function which relates transducer motion (or equivalently
voltage) to the applied force~\cite{morse_prd99}. We note that the overall calibration is
contained in the transfer function.  Defining a new transfer function (strain to transducer
motion)
\[ G_{f}(\omega) = -\frac{1}{2} \mu L_{e}\, \omega^{2}\, G(\omega) \] 
we write the signal as it appears in the data as simply
\begin{equation}
\label{siggy} 
H(\omega) = G_{f}(\omega) \, h_{bar}(\omega) 
\end{equation}
with $h_{bar}(\omega)$ given by the Fourier transform of Eq.~\ref{waveform3}.

At this point in the signal chain, in the interest of limiting the bandwidth of sample data
required, the signal is mixed with a reference (whose frequency is chosen to be between the
two detection mode frequencies) and low-pass filtered using a commercial lock-in
amplifier~\cite{mauceli_prd96}.  The lock-in provides both in-phase $(x)$ and
quadrature $(y)$ outputs, so both amplitude and phase information is available on a bandwidth of
125 Hz containing the resonant modes of the detector.  In software, the in-phase and
quadrature components were combined to form a complex data stream
\[ z(t) = x(t) + j\, y(t) \]
which was then demodulted to a 1 Hz bandwidth around each of the resonant
modes~\cite{mauceli_phd97}.  We describe the complete demodulation from 1 kHz to 1 Hz,
including both the hardware and software lockins, as  mixing the signal with a single
reference 
$ \exp \, [\, j (\omega_{r} t + \phi_{r})\, ]$, 
where $\omega_{r} $ is the reference frequency and $\phi_{r}$ is an unknown reference phase
(this phase can in fact be measured, but for the data set in question was not). Returning
to the time domain for clarity the mixed signal is
\begin{equation}
H(t) = [G_{f}(t) \star h_{bar}(t)]\exp \, [\, j (\omega_{r} t + \phi_{r})]\,
\end{equation}
where `$\star$' indicates a convolution. Using the expression from Eq.~\ref{waveform3}
we get

\begin{eqnarray}
\label{transout} 
H(t) &=& G_{f}(t) \star  ( \frac{1}{2} h_{c}\,(1+\cos^{2} i )\,[ \cos 2 \psi_{0} F^{*}_{+}(t) +
\sin 2 \psi_{0} F^{*}_{\times}(t)] \,\exp [-j (\Phi_{0}-\phi_{r}+(\omega_{0}-\omega_{r}) t)] \\
&+& \frac{j}{2} \, h_{c}\, (2\cos i) \, [\cos 2 \psi_{0} \, F^{*}_{\times}(t) - 
\sin 2 \psi_{0} \, F^{*}_{+}(t) ] \,
\exp [-j (\Phi_{0}-\phi_{r}+(\omega_{0}-\omega_{r}) t)] ) \nonumber \, .
\end{eqnarray}

Since $\omega_{r}$ is on the order of 1 kHz, the effect of the low-pass filtering
is to remove terms from $H(t)$ which contain the sum
frequencies $(\omega_{r}+\omega_{0})$.  Now returning to the frequency domain,
the signal, after demodulation, can be written
\begin{eqnarray}
\label{waveformd} 
H(\omega) &=& \frac{1}{2} h_{c}\,(1+\cos^{2} i )\, \exp [-j (\Phi_{0} - \phi_{r})]
\,G_{f}(\omega) \, [ \cos 2 \psi_{0} F_{+}(\omega^{'}) + \sin 2 \psi_{0}
F_{\times}(\omega^{'})] \\ 
&-& \frac{j}{2}\, h_{c}\, (2\cos i) \, \exp [-j (\Phi_{0} - \phi_{r})] \,G_{f}(\omega) \, 
[\cos 2 \psi_{0} \,F_{\times}(\omega^{'})-\sin 2
\psi_{0}\,F_{+}(\omega^{'}) ] \nonumber 
\end{eqnarray}
where $\omega^{'} = \omega \pm (\omega_{0}-\omega_{r})$ is the (positive or negative)
downconverted signal frequency.

\section{detection considerations}
\label{detecon}

Equation~\ref{waveformd} gives the form of the CW signal as it would appear in the Allegro data. 
We now ask: ``Is this signal in the data?"  Since the strength of the signal is small compared
to the detector noise (otherwise we would see it on a spectrum analyzer!), some work needs to
be done to answer this question.  We will use the standard maximum likelihood method to guide the
analysis, but in the end the question of whether a signal was detected or not will be answered
experimentally.

The likelihood function can be written 
\begin{equation}
	\Lambda = \frac{P(z|H)}{P(z|0)}
\end{equation} with 
$P(H|z)$ the probability that, given the observed data $z(t)$, the signal $H(t)$  is present,
$P(z|H)$ the probability that, given the signal $H(t)$ is present, we observed the data $z(t)$,
and $P(z|0)$ the probability that, given no signal present, we observe the data $z(t)$.

From~\cite{wan_zub}, if
\[ z = z_{1},z_{2},...,z_{N} \] 
is the time-ordered sequence of detector output (in our case imaginary) and
\[ H =  H_{1},H_{2},...,H_{N} \] 
is the signal waveform at the same sample times, then
\begin{equation} P(z|0) = \frac{1}{[ (4\pi)^{N} det||R_{gh}|| ]}
				       \exp (-\frac{1}{2} \sum_{g,h=0}^{N-1}R^{-1}_{gh}z_{g}z^{*}_{h})
\end{equation} 
and
\begin{equation}
 P(z|H) = \frac{1}{[ (4\pi)^{N} det||R_{gh}|| ]} 
 \exp \left (-\frac{1}{2} \sum_{g,h=0}^{N-1}R^{-1}_{gh}(z_{g}-H_{g})(z_{h}-H_{h})^{*} \right)
\end{equation} 
where the $R^{-1}_{gh}$ is the inverse of the autocorrelation matrix.

Substituting from above, the likelihood ratio is
\begin{equation}
    \Lambda = \exp (-\frac{1}{2} \sum_{g,h=0}^{N-1}R^{-1}_{gh}
       (-z_{g} H^{*}_{h} - H_{g} z^{*}_{h} + H_{g}H_{h}) )
\end{equation} 
which, given the properties of the complex autocorrelation matrix~\cite{mauceli_phd97}, can be
written
\begin{equation}
\Lambda = \exp \left( \Re \left \{ \, \sum_{g,h=0}^{N-1}R^{-1}_{gh}z_{g}H^{*}_{h}\,\right \} -
\frac{1}{2} \sum_{g,h=0}^{N-1}R^{-1}_{gh}H_{g}H^{*}_{h} \right ) 
\end{equation} where $\Re$ means ``take the real part''.

We define the following notation:
$x_{k}$ for the discrete Fourier transform of $x(t)$, evaluated at
$\omega_{k}=2 \pi k/ N \Delta t$, and $ Sn^{(1)}_{k}$ is the one-sided PSD. Using the
relation~\cite{finn_prd92}
\begin{equation}
\sum_{g,h=0}^{N-1}R^{-1}_{gh}x_{g}x^{*}_{h} = 2\Delta f \sum_{k=0}^{N-1}
    \frac{ x_{k} x^{*}_{k}}{ Sn^{(1)}_{k}}
\end{equation} 
and taking the natural logarithm of the the likelihood function (to remove the
exponential), we have 
\begin{equation}
\ln \Lambda = 2\Delta f \, \Re \left \{\, \sum_{k=0}^{N-1}
\frac{z_{k} H^{*}_{k}}{Sn^{(1)}_{k}} \,\right \} - \Delta f \sum_{k=0}^{N-1}
\frac{|H_{k}|^{2}}{Sn^{(1)}_{k}} \, .
\label{frsam}
\end{equation}

Substituting for $H$ gives,
\begin{eqnarray} \label{halfway2}
\ln \Lambda &=& 2\Delta f \, \Re \,(\,\exp [-j(\Phi_{0} + \phi_{r}) ]  \sum_{k=0}^{N-1}
\frac{ z_{k} \, G^{*}_{f_{k}}}{ Sn^{(1)}_{k} } \\ 
& \times & [ \frac{1}{2} h_{c}\,(1+\cos^{2} i )\, \cos 2 \psi_{0} \, F^{*}_{+_{k}} 
+ \frac{1}{2} h_{c}\,(1+\cos^{2} i )\, \sin 2 \psi_{0} \, F^{*}_{\times_{k}} \nonumber \\
&-& \frac{j}{2}\, h_{c}\, (2\cos i) \,\cos 2 \psi_{0} \, F^{*}_{\times_{k}}
+ \frac{j}{2}\, h_{c}\, (2\cos i) \,\sin 2 \psi_{0} \, F^{*}_{+_{k}} ]  ) \nonumber \\
&-& \Delta f \,\sum_{k=0}^{N-1} \frac{1}{ Sn^{(1)}_{k} } \nonumber \\
&\times & [ \frac{1}{4} h^{2}_{c}\,(1+\cos^{2} i )^{2} \, \cos^{2} 2 \psi_{0}
|G_{f_{k}}F_{+_{k}}|^{2} 
+ \frac{1}{4} h^{2}_{c}\,(1+\cos^{2} i )^{2} \, \sin^{2} 2 \psi_{0} 
|G_{f_{k}}F_{\times_{k}}|^{2} \nonumber \\
&+& \frac{1}{4} h^{2}_{c}\,(2\cos i )^{2} \, \cos^{2} 2 \psi_{0}
|G_{f_{k}}F_{\times_{k}}|^{2} 
+ \frac{1}{4} h^{2}_{c}\,(2\cos i )^{2} \, \sin^{2} 2 \psi_{0}
|G_{f_{k}}F_{+_{k}}|^{2} ] \nonumber
\end{eqnarray}
where terms of the form 
\[ \sum_{k=0}^{N-1} \, F_{+_{k}} \, F^{*}_{\times_{k}} 
\] 
sum to zero.

To simplify the notation of Eq.~\ref{halfway2}, we make the following definitions:
\begin{eqnarray} 
q_{+} &=& \Delta f \,\sum_{k=0}^{N-1} \frac{z_{k} \, G^{*}_{f_{k}}
F^{*}_{+_{k}}}{Sn^{(1)}_{k}} \\
q_{\times} &=& \Delta f \, \sum_{k=0}^{N-1} \frac{z_{k} \, G^{*}_{f_{k}}
F^{*}_{\times_{k}}}{Sn^{(1)}_{k}} \nonumber \\
\rho_{+} &=& \Delta f \,\sum_{k=0}^{N-1} \frac{|G_{f_{k}}\,
F_{+_{k}}|^{2}}{Sn^{(1)}_{k}} \nonumber \\
\rho_{\times} &=& \Delta f \,\sum_{k=0}^{N-1} \frac{|G_{f_{k}}\,
F_{\times_{k}}|^{2}}{Sn^{(1)}_{k}} \nonumber \\ 
a_{1} &=&  h_{c} (1+\cos^{2} i) \cos 2 \psi_{0} \nonumber \\
a_{2} &=&  h_{c} (1+\cos^{2} i) \sin 2 \psi_{0}  \nonumber \\
a_{3} &=&  h_{c} (2 \, \cos i) \cos 2 \psi_{0} \nonumber \\
a_{4} &=&  h_{c} (2 \, \cos i) \sin 2 \psi_{0}  \nonumber \, .
\end{eqnarray}

The four quantities $\{ q_{+}, q_{\times}, \rho_{+}, \rho_{\times} \}$ are familiar from signal
processing theory.  The first two, $\{ q_{+}, q_{\times}
\}$, are the outputs from applying independent optimal filters in the frequency domain for the two
polarizations of the gravity wave.  The second pair, $\{ \rho_{+}, \rho_{\times} \}$ are the signal to
noise ratios (energy) for the two polarizations. These four terms are completely specified, up to the
unknown signal frequency.  The four $a_{i}$'s are the ``amplitudes'' containing the desired
astrophysical information.

Substituting these into the likelihood function,
\begin{eqnarray}\label{liklhd2}
\ln \Lambda &=&  a_{1} \Re \left \{ \, \exp [j(\Phi_{0} - \phi_{r}) ] \, q_{+} \, 
\right \} 
+ a_{2} \Re \left \{ \, \exp [j(\Phi_{0} - \phi_{r}) ] \, q_{\times} \, 
\right \} \nonumber \\
&-&  a_{3} \Re \left \{ \, j \,  \exp [j(\Phi_{0} - \phi_{r}) ] \, q_{\times} \, 
\right \}  
+  a_{4} \Re \left \{ \, j \, \exp [j(\Phi_{0} - \phi_{r}) ] \, q_{+} \, 
\right \} \nonumber \\
&-& \frac{1}{4} ( a_{1}^2 \, \rho_{+} + a_{2}^2 \, \rho_{\times} + a_{3}^2 \, \rho_{\times} +
a_{4}^2 \, \rho_{+} )\, .\nonumber
\end{eqnarray}

We maximize the likelihood function over the $a_{i}$'s which yields the following four expressions:
\begin{eqnarray}\label{ays}
a_{1} &=&  2 \frac{|q_{+}|}{\rho_{+}} \cos (\Phi_{0} - \phi_{r} + \phi_{+}) \\
a_{2} &=& 2 \frac{|q_{\times}|} {\rho_{\times}} \cos (\Phi_{0} - \phi_{r} + \phi_{\times}) 
\nonumber \\
a_{3} &=& -2 \frac{|q_{\times}|}{\rho_{\times}} \sin (\Phi_{0} - \phi_{r} + \phi_{\times})
\nonumber \\
a_{4} &=& 2 \frac{|q_{+}|}{\rho_{+}} \sin (\Phi_{0} + \phi_{r} - \phi_{+})\nonumber  
\end{eqnarray}
where the filtered ouputs have been expressed as
\[
q_{(+,\times)} = |q_{(+,\times)}| \exp (j \phi_{(+,\times)}) 
\]
with $\,  \phi_{(+,\times)} = \arg [ q_{(+,\times)} ] $. 
Finally, 
\begin{equation}\label{energ}
 \sum_{i=1}^{4} a^{2}_{i} = 4(\frac{|q_{+}|^{2}}{\rho_{+}^{2}} + 
\frac{|q_{\times}|^{2}}{\rho_{\times}^{2}}) = h^{2}_{c}\,[(1+\cos^{2} i)^{2} +  4\cos^{2} i] 
= \left <  h^{2}_{+} + h^{2}_{\times} \right >
\end{equation}
which, to factors, is the energy of the gravitational wave.  Note that this
depends on the unknown signal frequency $f_{0}$.  As described below, we will assume values for
the gravitational wave frequency, and use that to calculate Eq.~\ref{energ}.  The end result of the
analysis is a ``spectrum'' of energy at a given signal frequency.  We shall report this as a
``strain amplitude'', given by the square root of Eq.~\ref{energ}
\[  h_{s}(f_{0}) = \sqrt{ h^{2}_{+}(f_{0}) + h^{2}_{\times}(f_{0}) } \, .\]

\section{experimental considerations}
\label{expcon}

Calculation of Eq.~\ref{energ} involved a number of steps.  First, the archived data was read
off tape and narrowbanded to a 1 Hz bandwidth around each of the resonant
modes~\cite{mauceli_phd97} where ALLEGRO is most sensitive.  This was done to reduce the
computational requirements.  Next, the mode amplitudes were ``cleaned'' of large, transient
events (section~\ref{datagap}).  Finally, the cleaned data was discrete Fourier transformed and
optimal filters were applied (section~\ref{opfilt}).

\subsection{data selection}
\label{datagap}

Even with ALLEGRO's high duty cycle, there were periods of missing or unusable data.   Data
losses came in basically three flavors.  (1) Transient electronic effects which lasted on the
order of a second.  (2)  Longer periods when the detector was undergoing some form of
maintenance.  (3)  The clock losing phase lock to WWVB.  The transient disturbances were the
most frequent, occurring at a rate of one or two per day.  They usually involved a sudden
change in the flux threading the SQUID loop (hence the name ``flux jumps'').  The majority of
the flux jumps occurred when the dc level of the SQUID reached a pre-determined maximum (5
volts).  The electronics controlling the SQUID then reset the dc voltage to zero, causing a
short and violent jump in the in-phase and quadrature channels of the data, as shown in
Fig.~\ref{spike}.

Frequently electronic interference reaching the SQUID caused flux jumps.  In the past when data
tapes were erased (the degaussing takes place in a separate room from the main experiment), the
end of the degaussing cycle produced a noise spike which traveled through the wiring in the
wall, through the computers, and from there to the SQUID.  Once recognized, an isolating
transformer was placed between the degausser and the wall socket, fixing the problem.  Another
common type of transient signal is a ``spike''.  These look similar to flux jumps in the data
and there is some suspicion that they are in fact flux jumps, but essentially they are of
unknown origin.  All of these noisy periods were short enough so that the affected data could
be removed and the resulting gap interpolated across.  This was done using the MATLAB {\em
interp1} routine.  Interpolation was performed on the resampled data.  It was usual to
interpolate across 1-5 seconds of data, using the 10-20 seconds of data before and after the
gap for the interpolation template. 

For longer sections of unusable data or for periods of missing data, the analysis was stopped
and restarted after the disturbance.  By ``restarted'' we mean that the accumulation of data
was stopped, the accumulated data purged, and the analysis started up again on the data
immediately following the disturbance. The most common cause of data loss was transferring
liquid helium into the dewar which removed a couple of hours of data every week.  Another cause
of long stretches of unusable data were large excitations of the resonant modes due to
earthquakes around the globe.  Earthquakes were identified by a unique signature in the low
frequency housekeeping channel. It was usual for an earthquake to produce multiple large
excitations over a few tens of minutes, often resulting in saturation of the A/D's.   Computer
down time, calibration of the detector and other maintenance all caused gaps in the data,
although infrequently.  

The final type of data loss was associated with the WWVB clock.  Frequently when the weather
between Baton Rouge and NIST at Boulder, Colorado was bad, the clock we used to control
the sampling of the data lost phase lock to the WWVB radio signal.  When this happened, the
clock's internal oscillator ``freewheeled'' with the result that the time between samples was
no longer consistently 8 ms.  Deviations in the sampling rate from  8 ms were called ``timing
jumps''.   A jump in the time between when samples were taken  produced a corresponding jump in
the phase of modes and the calibrator signal as shown in Fig.~\ref{tjump}.  The most frequent
jumps were on the order of 1-2 ms, producing a phase jump in a sinusoidal signal at the mode
frequencies of approximately 1/100 of a cycle.  This was considered an acceptable level of
uncertainty in the ability to track the phase of a potential gravity wave signal. These small
jumps were noted but ignored.  Larger jumps produced correspondingly larger jumps in phase and
were considered unacceptable.  When they occurred, the analysis was stopped at the timing jump
and restarted again after the glitch.  The frequency and size of the timing jumps were highly
variable.  During the winter of 1994 the smaller jumps occurred almost once per day while the
larger jumps rarely happened.  By the spring of that year, the trend was reversed and much data
was lost due to the inconsistency of the clock.  

\subsection{the filters}
\label{opfilt}

The ability of our analysis filters to match the phase of the gravity wave determined the 
amount of continuous data which could be analyzed at one time.  We refer to this as a
``record''.  Ideally, the length of a record would be set by the available computational
facilities.  This was not the case.  Instead, the record length was limited by our electronics. 
The function generator which supplied the reference frequency to the lock-in detector had some
drift associated with it.  This drift limits the ability of the filter to match the phase of
the gravity wave signal in the data.   Careful measurements of the drift led to the
conservative conclusion that roughly 28~hours ($10^5$~sec) was the longest period of time for
which we could expect the filter to remain in phase with the signal~\cite{mauceli_phd97}. The
less conservative conclusion was roughly three times longer, but we prefer to be cautious with
the analysis.  The reference signal to the lock-in has since been phase locked to GPS, greatly
improving the phase stability.  Given the above considerations, there were 34 records
available from days 1-94 of 1994, for a total of 944.44 hours of data.

Performing a discrete Fourier transform on $10^5$~sec worth of data results in a
frequency resolution of the search of $10^{-5}$ Hz.  The optimal filters also depend on the
signal frequency.  With no known source available, we assume a there is a potential signal
every 10 $\mu$Hz in the ranges 896.30-897.30 Hz and 919.76-920.76 Hz.  Each assumed
signal frequency was used in turn to create the $F_{+}$ and $F_{\times}$ components of the
signal template.  

The other two components of the filters, the bar transfer function and the power spectral
density, depend on the resonant frequency and damping time of the coupled bar-transducer
system.  As both quantities experienced slow drifts due to, for example, temperature changes in
the dewar, it was necessary to measure them for each record (see Fig.~\ref{resfreq} and
Fig.~\ref{damptime} respectively).  For each mode, a low variance power spectral density was formed
from the cleaned data.  A Lorenzian curve was then fit to each PSD using the MATLAB {\em curvefit}
routine, which finds the best fit to a function in the least-squares sense.  The Lorentzian was
characterized by four parameters: the white-noise level ($S_{0}$), the peak height ($S_{1}$), the
frequency of the resonance ($\omega_{\pm}$),  and the width (or the damping time - $\tau_{\pm}$).  The
resulting fitted parameters for each record were then stored on disk to be retrieved as necessary.  
The power spectral density component of the optimal filter was calculated for a record by
\begin{equation} Sn(\omega_{k}) = S_{0} + \frac{(S_{1}-S_{0})}{\tau^{2}_{\pm}} [
(\omega^{2}_{k}-\omega^{2}_{\pm})^{2} + \omega^{2}_{k}/\tau^{2}_{\pm}]^{-1} \, .
\end{equation}
Similarly, the bar transfer function was calculated from the functional
\begin{equation} G(\omega_{k}) = k_{\pm}\, \frac{\omega_{\pm}}{\tau_{\pm}}\, [\omega^{2}_{\pm} +
j\, \omega_{k}/\tau_{\pm} - \omega^{2}_{k}]^{-1}
\end{equation}
where the parameters are those given above.  The constant $k_{\pm}$ sets the calibration
for each mode~\cite{morse_prd99}.

\subsection{computational reduction}

Even though this search was directed towards only a small section of the sky, it was still cpu
intensive on the available computing facilities.  Applying $2\times 10^{5}$ different filters,
where each new set of filter coefficients required a Fourier transform on a time sequence of
$10^{5}$ elements, was prohibitively time consuming. 

The signal frequency enters into calculation of the filter coefficients in two places;
through the carrier wave and in the Doppler shift.  However, as the phase modulation is
slowly varying with respect to the carrier wave, it was not necessary to re-calculate the phase
correction due to the Doppler shift for each assumed signal frequency.  Instead, the Doppler
shift was calculated at specific choices of the signal frequency, which were then used to approximate
signal templates over a range of frequencies. 

The maximum fractional frequency shift of a signal from the two sources over the course of a year is
\[
  \frac{|\delta f (f_{0})|}{f_{0}} \sim \left \{ \begin{array}{ll}
			   5 \times 10^{-5} & \mbox{ 47 Tuc} \\
	     10^{-4}          & \mbox{ galactic center}
      \end{array}
\right .\]
at the signal frequencies of interest.
To approximate all phase shifts in a range of signal frequencies $f_{0} \pm \Delta f_{0}$
with a calculation at a single signal frequency, then the error made must be less than the
frequency resolution of the search.  Defining the error as
\[
  \delta f (f_{0} \pm \Delta f_{0}) - \delta f (f_{0}) = 5\times 10^{-5} \Delta f_{0} 
\] 
we have the maximum range of signal frequencies over which the approximation is valid is
given by
\[ 
  |\,\Delta f_{0}\,| \leq \left \{ \begin{array}{ll}
  0.2 \: & \mbox{Hz, 47 Tuc} \\
	 0.1 \: & \mbox{Hz, galactic center}
	  \end{array}
\right .
\] 
Using this information, we calculated independent $F_{+},F_{\times}$ components of the signal template
at the frequencies 896.45 Hz, 896.80 Hz, 897.15 Hz, 919.91 Hz, 920.26 Hz and 920.61 Hz for the 47 Tuc
analysis.  For the galactic center analysis, signal templates were calculated every 0.2 Hz in the
range  896.40 Hz - 897.20 Hz for the minus mode, and 919.86 Hz - 920.66 Hz for the plus mode.

\section{results}
\label{results}

The real and imaginary components of the data stream, in the absence of a signal, are zero mean
Gaussian distributed.  Neither the Fourier transform or the filtering changes the underlying
distribution and therefore the real and imaginary parts of the filtered outputs $q_{(+,\times)}$ are
also individually zero mean Gaussian distributed.  Given this, calculation of Eq.~\ref{energ} 
at a particular choice of the signal frequency, taken from a particular data record, results in
a sample drawn from a chi-squared distribution with four degrees
of freedom~\cite{whalen}.  If the signal is absent, this is
given by
\begin{equation} p(E) = \frac{1}{(2 \sigma^2)^2} E \, \exp\, ( -E / 2\sigma^2 ) \, .
\end{equation}
where $E = h^{2}_{s}(f_{0})$. 

The total number of available data records provide an ensemble of experiments each making
$2\times 10^{5}$ measures of $E$ at the assumed values for
$f_{0}$.  Since the signal is expected to be long-lived, much longer than the operational time
of the detector, we improve our estimate of the energy by taking the ensemble average of the
individual measures:
\begin{equation}
\bar{E} = \frac{1}{N}\sum_{i=1}^{N} E_{i}
\end{equation} 
where $N=34$ is the number of available data records.  We calculate the strain amplitude 
$h_{s}(f_{0})$ by taking the square root of $\bar{E}$.  As each ensemble is generated by the sum of
four terms (Eq.~\ref{energ}) and there are 34 data records averaged together, from the central limit
theorm we can accurately describe the distribution of $h_{s}(f_{0})$ as gaussian.

The resulting ``spectra'' of $h_{s}(f_{0})$ vs. $f_{0}$ are shown in
Fig.~\ref{tucp} for 47 Tuc and
Fig.~\ref{gcp} for the galactic center analysis.  It is
important to remember that the abscissa is not a Fourier frequency but rather the assumed
frequency of the gravitational wave.  That the plus mode shows a lower strain value than the
minus mode reflects the fact that the plus mode has a higher quality factor and is therefore
more sensitive.  That the results of the galactic center analysis seem somewhat ``shifted'' with
respect to the 47 Tuc results is due to the effects of the Doppler shifts.  For a signal from 47 Tuc
during the period over which data was analyzed, a signal at a particular frequency, for example 920
Hz, experienced periods of both red shift and blue shift.  On average, therefore, the signal was
experiencing detector noise at roughly 920 Hz.  For the galactic center analysis, during this period
an incoming signal was experiencing maximum blue-shift, and was never red-shifted.  Therefore, a
signal at 920 Hz would be experiencing detector noise at a slightly higher frequency, resulting in a
graph that seems ``a bit skewed to the right'' with respect to the 47 Tuc analysis.

The question still remains if there is a real signal at a particular frequency.  It is unlikely
that the detector is dominated by CW radiation.  We expect the number of real signals which
would be extracted by our analysis to be small with respect to the number of possible signals
(i.e. the number of frequencies at which we assume there to be a signal).  We can therefore
answer the question experimentally.  We normalize
$h_{s}(f_{0})$ at each $f_{0}$ to its mean value.   This
new variable has a distribution independent of the detector noise at each signal frequency.

The histograms of the resulting normalized spectra are shown in Fig.~\ref{tucnorm} for 47 Tuc and
Fig.~\ref{gcnorm} for the galactic center analysis.  Given our reasonable assumption that
the detector is not wave-dominated, the majority of the normalized measures form an
experimental estimation of the parent distribution for the detector noise from which each
individual measure is drawn.  If the strain amplitude at a particular signal frequency lies
outside this distribution, it is identified as a candidate for a real signal.

\section{discussion}
\label{discu}

Neither Fig.~\ref{tucnorm} or Fig.~\ref{gcnorm} shows any evidence for ``outliers'' which would
suggest the possibility of having observed gravitational radiation from pulsar spin-down.  
The existence of burst-like non-gaussian noise in resonant bar data is well known, and this noise must
be dealt with in any search for burst-like signals of gravitational waves. 
The encouraging result of this analysis is that for CW signals 
there is not a serious problem with non-gaussian noise in the
resonant bar data. Thus, at least at the level of the current work, all the noise sources are
well understood and the overall sensitivity may be improved in the future simply by inceasing the
length of the data record. 
We note that care was taken to use only data when the
detector was operating well and procedures were implemeted to smooth out other non-stationary
effects.  Astrophysically, at this level of sensitivity, the
actual spin-down rate of a pulsar due to the energy lost as gravitationl radiation 
would be large enough that the signal and its filter template would
go out of phase by half a cycle after of order 5000 s, much less than the length of one record.  This
effect can be taken into account by including one or more spin-down parameters in the signal
template, at the cost of significantly increasing the computational time of the search.  Since there
was no known source for ALLEGRO and somewhat limited computing resources available, we have
essentially made a demonstration of the capabilities of resonant detectors for this type of search. 

However, it is important to note that even with these restrictions, the analysis has reached a level
of sensitivity which is astrophysically interesting.  If a source were to 
become known with the correct frequency, and, as our sensitivity was limited by hardware which has been
substantially improved, it seems likely that such a signal could be detected and new astrophysics
learned.

\section{conclusion}

We have searched for sources of continuous gravitational radiation from the globular
cluster 47 Tuc and the galactic center at signal frequencies near 1 kHz using data taken by the
ALLEGRO gravitational radiation detector in the first three months of 1994.  No candidate signals were 
found, and a constraint of $8\times 10^{-24}$ was put on gravitational radiation emitted from pulsar
spin-down at both locations.

\acknowledgments{The authors thank Sam Finn and Joel Tohline for many helpful discussions.  
Much of the analysis and the final writing of this paper was done at the INFN Laboratori
Nazionali di Frascati.  E.M. thanks G. Pizzella, E. Coccia, and the ROG collaboration for their
support.  The LSU group was supported by the National Science Foundation under Grant No.
PHY-9311731.}

\bibliographystyle{prsty}

\begin{figure}
	\caption{Frequency shift of a signal from 47 Tuc arriving at the
   ALLEGRO detector on January 1-2, 1994. }
	\label{fshift}
\end{figure}

\begin{figure}
	\caption{The reception patterns for the plus polarization (solid line) and the cross
   polarization (dot-dashed line) of a signal from 47 Tuc. }
	\label{recpat}
\end{figure}

\begin{figure}
	\caption{A SQUID reset in the in-phase and quadrature channels of the data stream.}
	\label{spike}
\end{figure}

\begin{figure}
	\caption{A 1 ms jump in the sampling time as it appears in the continuous system test signal (CST-a
sinusoidal driving force applied to the bar at 865 Hz) \protect\cite{mauceli_prd96}.}
	\label{tjump}
\end{figure}

\begin{figure}
  \caption{The measured resonant frequencies for each mode and each data record.  The freqeuncy is
given as a shift relative to the measured frequency for the first data record.}
		\label{resfreq}
\end{figure}

\begin{figure}
  \caption{The measured damping times for each mode and each record.}
		\label{damptime}
\end{figure}

\begin{figure}
	\caption{The strain amplitude of a gravitational wave from 47 Tuc at each assumed signal
          frequency near the plus and minus resonant
          modes.  The lower values of the strains of the plus modes relative to the minus mode are due to the higher mechanical Q of the plus mode.  }
	\label{tucp}
\end{figure}

\begin{figure}
	\caption{The strain amplitude of a gravitational wave from the galactic center at each
     assumed signal frequency near the plus and minus resonant modes. }
	\label{gcp}
\end{figure}

\begin{figure}
	\caption{Histograms of the normalized ``spectrum'' for signals from 47 Tuc near the minus and
          plus modes.  The solid curve is a gaussian distribution with mean of one and variance derived from the normalized spectrum.}
	\label{tucnorm}
\end{figure}

\begin{figure}
	\caption{Histograms of the normalized ``spectrum'' for signals from the galactic center
 near the minus and plus modes. The solid curve is a gaussian distribution with mean of one and variance derived from the normalized spectrum.}
	\label{gcnorm}
\end{figure}

\begin{table}
  \caption{Coordinates for the two source locations}
   \protect \label{tab:coords}
  \centering \begin{tabular}{|c|c|c|c|} 
     Source & RA (hrs:mins:secs) & declination (degs:mins:secs) & distance (kpc) \\ 
          \hline
     galactic center & 17:42:29.3 & -28:59:18 & 8.0  \\ \hline
     47 Tuc & 00:24:06  & -72:04:00  & 4.5  \\ 
   \end{tabular}
 \end{table}  

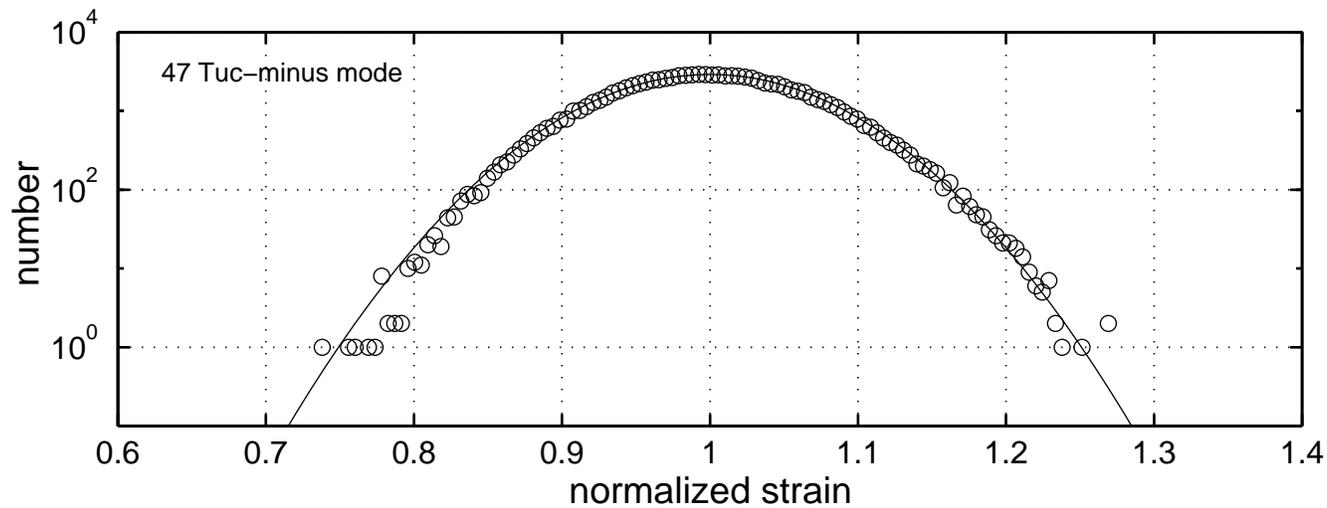
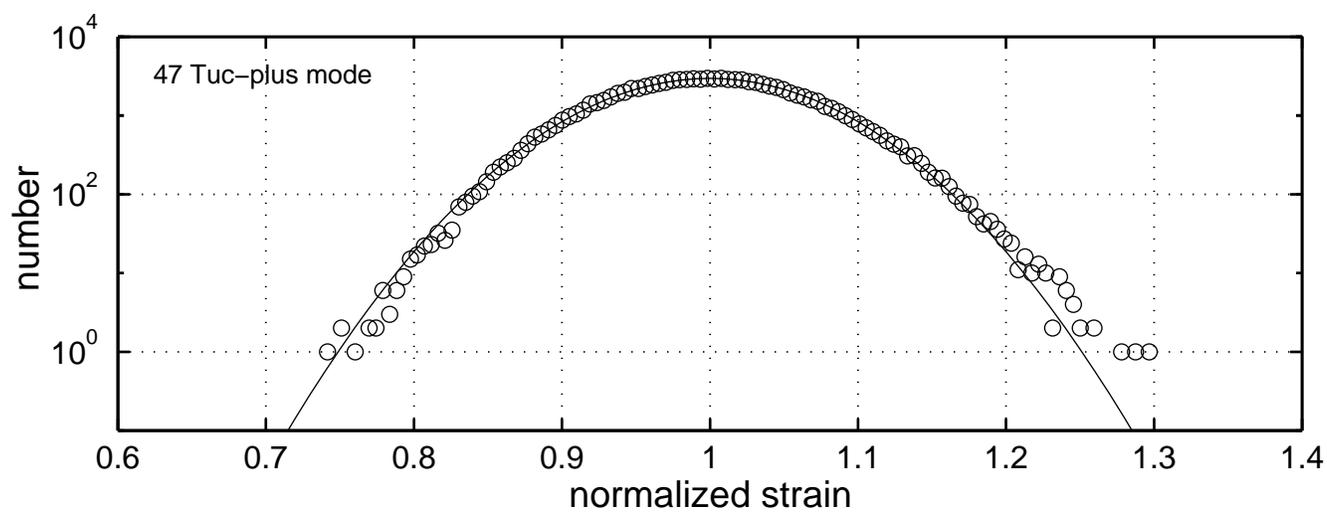
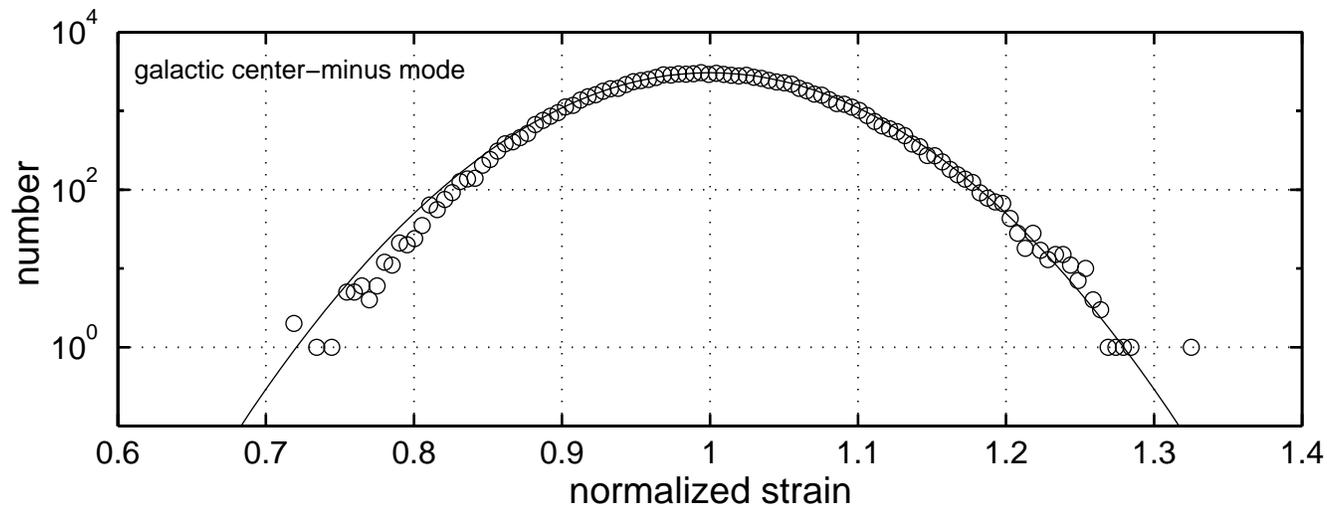
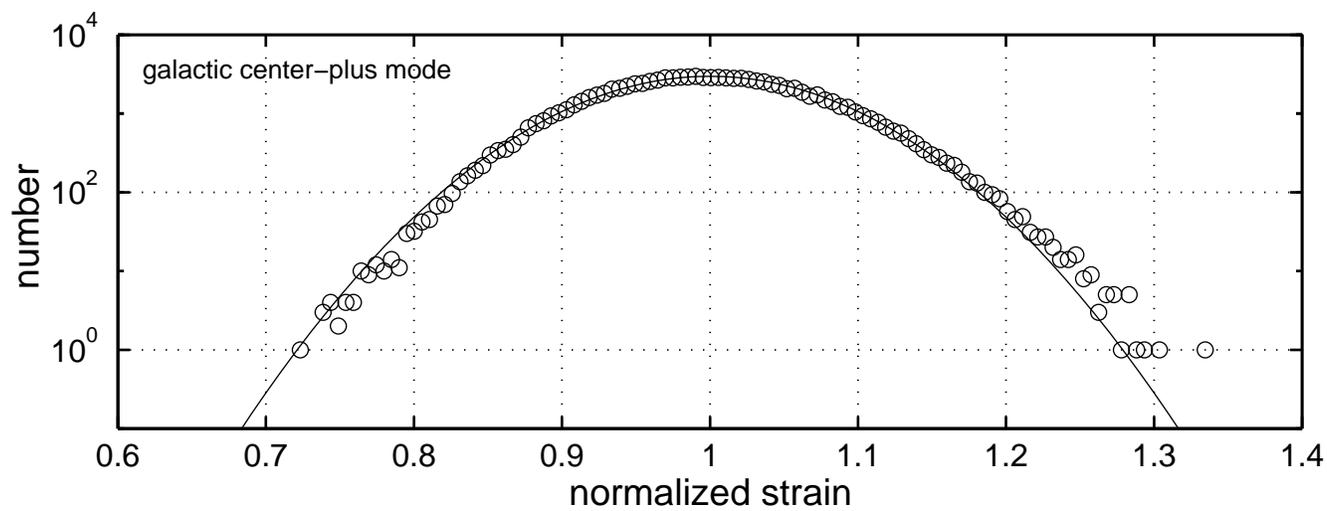
\end{document}